\documentclass[11pt,a4paper]{article}
\pdfoutput=1 

\usepackage{jcappub} 

\usepackage[T1]{fontenc} 
\usepackage{textcomp}
\usepackage{gensymb}
\usepackage{multirow}
\usepackage{lineno}
\usepackage{rotating}
\usepackage{makecell}
\usepackage{epstopdf}
\usepackage{soul}
\title{\boldmath Searches for neutrinos from cosmic-ray interactions in the Sun using seven years of IceCube data}
\author[15]{M. G. Aartsen,}
\author[54]{M. Ackermann,}
\author[15]{J. Adams,}
\author[11]{J. A. Aguilar,}
\author[19]{M. Ahlers,}
\author[45]{M. Ahrens,}
\author[25]{C. Alispach,}
\author[36]{K. Andeen,}
\author[51]{T. Anderson,}
\author[11]{I. Ansseau,}
% ...
\affiliation[11]{Universit{\'e} Libre de Bruxelles, Science Faculty CP230, B-1050 Brussels, Belgium}
% ...
\affiliation[15]{Dept. of Physics and Astronomy, University of Canterbury, Private Bag 4800, Christchurch, New Zealand}
% ...
\affiliation[19]{Niels Bohr Institute, University of Copenhagen, DK-2100 Copenhagen, Denmark}
% ...
\affiliation[25]{D{\'e}partement de physique nucl{\'e}aire et corpusculaire, Universit{\'e} de Gen{\`e}ve, CH-1211 Gen{\`e}ve, Switzerland}
% ...
\affiliation[36]{Department of Physics, Marquette University, Milwaukee, WI, 53201, USA}
% ...
\affiliation[45]{Oskar Klein Centre and Dept. of Physics, Stockholm University, SE-10691 Stockholm, Sweden}
% ...
\affiliation[51]{Dept. of Physics, Pennsylvania State University, University Park, PA 16802, USA}
% ...
\affiliation[54]{DESY, D-15738 Zeuthen, Germany}
% ...

\emailAdd{analysis@icecube.wisc.edu}

\abstract{Cosmic-ray interactions with the solar atmosphere are expected to produce particle showers which in turn produce neutrinos from weak decays of mesons. These solar atmospheric neutrinos (SA$\nu$s) have never been observed experimentally. A detection would be an important step in understanding cosmic-ray propagation in the inner solar system and the dynamics of solar magnetic fields. SA$\nu$s also represent an irreducible background to solar dark matter searches and a detection would allow precise characterization of this background. Here, we present the first experimental search based on seven years of data collected from May 2010 to May 2017 in the austral winter with the IceCube Neutrino Observatory. An unbinned likelihood analysis is performed for events reconstructed within 5 degrees of the center of the Sun. No evidence for a SA$\nu$ flux is observed. After inclusion of systematic uncertainties, we set a 90\% upper limit of $1.02^{+0.20}_{-0.18}\cdot10^{-13}$~$\mathrm{GeV^{-1}cm^{-2}s^{-1}}$ at 1~\textrm{TeV}.}

\begin{document}
\maketitle
\flushbottom


\section{Introduction}
\label{sec:intro}
Neutrinos can be produced as a result of cosmic-ray interactions in the solar atmosphere. Cosmic rays interact with nuclei in the solar atmosphere, producing particle showers including pions and kaons. The decays of these mesons produce so called ``Solar Atmospheric Neutrinos'' (SA$\nu$s). Theoretical flux predictions of SA$\nu$s and detailed process discussions have been given in~\cite{Seckel1991,Moskalenko1991,Moskalenko1993,IT1996,Hettlage1999,Fogli2006,Edsjo:2017kjk,FJAWs2017}. The neutrino production process in the solar atmosphere is similar to that of the terrestrial atmospheric neutrinos, with the notable difference that mesons generated in the solar atmosphere tend to decay before they can re-interact or lose a significant fraction of their energy, due to the larger and thinner atmosphere. As a result, the neutrino spectrum from the solar atmosphere is expected to be harder compared to that from the Earth, where the spectrum is steepened due to interactions of the secondary mesons, see \textit{e.g.}~\cite{CombineH4A}. This difference makes the spectra distinguishable and is used as a main criteria in our search for the SA$\nu$ flux. 
A search for solar atmospheric neutrinos has never been experimentally performed and this work is the first of its kind. 

The production process of SA$\nu$s is closely connected to that of gamma-rays through the decays of neutral pions and other mesons.  Evidence for solar gamma rays was first reported in a re-analysis of EGRET data~\cite{EGRET2008}. Recently, the Fermi-LAT Collaboration reported the observation of a steady gamma-ray emission from the solar disk with energies up to 10~\textrm{GeV}~\cite{Fermi2011}. 

In addition to the solar disk emission predominantly due to neutral pion decays from cosmic-ray interactions in the solar atmosphere, an extended inverse Compton signal from cosmic-ray electron interactions with the solar photon field was also observed. A follow-up analysis on the solar disk emission based on six years of public Fermi-LAT data has shown that the energy spectrum extends beyond 100~\textrm{GeV} and anticorrelates with the solar activity~\cite{Kenny2016}. This was confirmed with an extended nine year analysis~\cite{Tim2018}. The magnetic field near the Sun is complex and strongly time-dependent. Gamma-ray production is expected to be significantly enhanced above 10~\textrm{GeV} in case of a more intense magnetic field. However, the effects on the neutrino production are found to be negligible~\cite{Mazziotta:2020uey}.
Further, the observed gamma-ray spectrum shows a potential dip~\cite{Qing2018} and points to an inhomogeneous emission between the equatorial plane and the polar region of the Sun~\cite{Tim2018}. Unexpectedly, the observed gamma-ray flux is about six times higher~\cite{Kenny2016, Tim2018} than theoretical predictions~\cite{Seckel1991}. The High Altitude Water Cherenkov (HAWC) gamma-ray observatory has searched for gamma rays beyond the energies accessible by Fermi-LAT. HAWC reported no evidence of \textrm{TeV} gamma-ray emission in three years data and has set flux bounds~\cite{HAWC2018}. The recent observation of gamma-ray emission from the Sun makes the search for solar atmospheric neutrinos very timely. The combined gamma-ray and neutrino data are expected to be vital to understand the solar atmospheric processes and cosmic-ray transport in the inner solar system~\cite{Seckel1991,BeckerTjus:2019rqu}.

IceCube is the world's largest neutrino telescope and is optimized to detect high-energy (TeV) neutrinos. IceCube's acceptance to high-energy neutrinos and sub-degree-scale angular resolution to muon neutrinos makes it ideally suited to search for SA$\nu$s at TeV scales where the flux of SA$\nu$s is expected to dominate over that from terrestrial atmospheric neutrino backgrounds. In our analysis we rely on well established event selection criteria~\cite{ICPS2016,ICRC2017_Sample} and a data set that has previously been used to study distant neutrino sources~\cite{ICPS2015,ICPS2016,ICPS2019}.

This paper is structured as follows: Section~\ref{sec:Detector} describes the IceCube detector. Predictions for signal energy spectra and backgrounds to this analysis are given in Section~\ref{sec:SigBkg}. The data samples and the simulations are described in Section~\ref{sec:DataSim}. Analysis method to search for SA$\nu$s and systematic uncertainties are given in Section~\ref{sec:Analysis}. The results are presented in Section~\ref{sec:Results}. Finally, Section~\ref{sec:Conclusion} presents our conclusions and we discuss the prospects for future analysis and its applications.

\section{The IceCube Neutrino Observatory}
\label{sec:Detector}
The IceCube Neutrino Observatory consists of the IceTop surface array~\cite{IceCube:2012nn} and the in-ice array~\cite{ICDesign2006} to detect Cherenkov light from relativistic charged particles, \textit{e.g.} muons and electrons produced by high-energy neutrino interactions. The in-ice array is installed in the Antarctic ice at depths between 1450~m to 2450~m with 5160~Digital Optical Modules (DOMs)~\cite{ICDesign2006}. The in-ice array is comprised of 86~vertical strings (IC86) arranged in an approximately hexagonal geometry, instrumenting a volume of 1~$\mathrm{km^{3}}$. Each DOM is made of a downward-pointing 10-inch photomultiplier tube (PMT)~\cite{ICPMT2010} to detect Cherenkov photons. The DOM includes readout electronics and a high-voltage power supply~\cite{ICDA2008}. The PMT and its electronics are protected by a spherical glass vessel. The optical properties of the ice have been studied and are used to build a detailed response model of the detector~\cite{ICLED2013}. This model includes depth-dependent scattering and absorption, optical anisotropy and tilt. This analysis uses data from the full array as well as one year of data from before IceCube construction was complete, when it consisted of 79~strings (IC79). Since 2010, IceCube has run stably with an average detector uptime greater than 99\%~\cite{ICJINST}.

\section{Signal and background predictions} 
\label{sec:SigBkg}

\subsection{Signal predictions}
\label{Subsec:SigExp}
The first theoretical calculations for SA$\nu$s date back to 1991~\cite{Seckel1991,Moskalenko1991}. The authors modeled gamma-ray, neutrino, antiproton, neutron, and antineutron fluxes that are initiated by the interactions of cosmic rays with the solar atmosphere. The flux originates from the solar disk as cosmic rays that are mirrored in the solar atmosphere are expected to contribute significantly to the flux~\cite{Seckel1991}. While these early predictions were based on semi-analytical calculations, full numerical simulations of the interactions based on the Monte Carlo method have been performed in Ref.~\cite{IT1996}. In more recent publications~\cite{Edsjo:2017kjk,FJAWs2017}, uncertainties in the predicted neutrino energy spectra from the choice of the primary cosmic-ray flux models, particle interaction models, solar density models, and neutrino oscillation parameters are discussed. 
Neutrinos generated in the atmosphere of a non-magnetic Sun are propagating through the Sun and their attenuation due to absorption is included. The impact of mirroring of charged particles in solar magnetic flux tubes anchored at the bottom of the photosphere is not considered as it is expected to be sub-dominant at energies above 100~GeV~\cite{Seckel1991}. 
We will discuss the energy dependent spatial distribution of the solar disk flux in Sec.~\ref{Subsubsec:SourceDist}. 

Solar atmospheric neutrino fluxes have been implemented in the simulation framework, \textsf{WIMPSim}~\cite{WIMPSim}, which we used for our signal prediction. We obtained neutrino fluxes for all neutrino flavor channels. However, for our analysis we only consider the $\nu_\mu+\bar{\nu}_\mu$ channel to benefit from IceCube's excellent angular resolution O(1\degree). The impact of additional flavor contributions are discussed as part of our systematic uncertainty studies (see Sec.~\ref{sec:DataSim} and~\ref{Subsec:Systematic}).

\begin{figure}[tbp]
\centering 
\includegraphics[width=.7\textwidth]{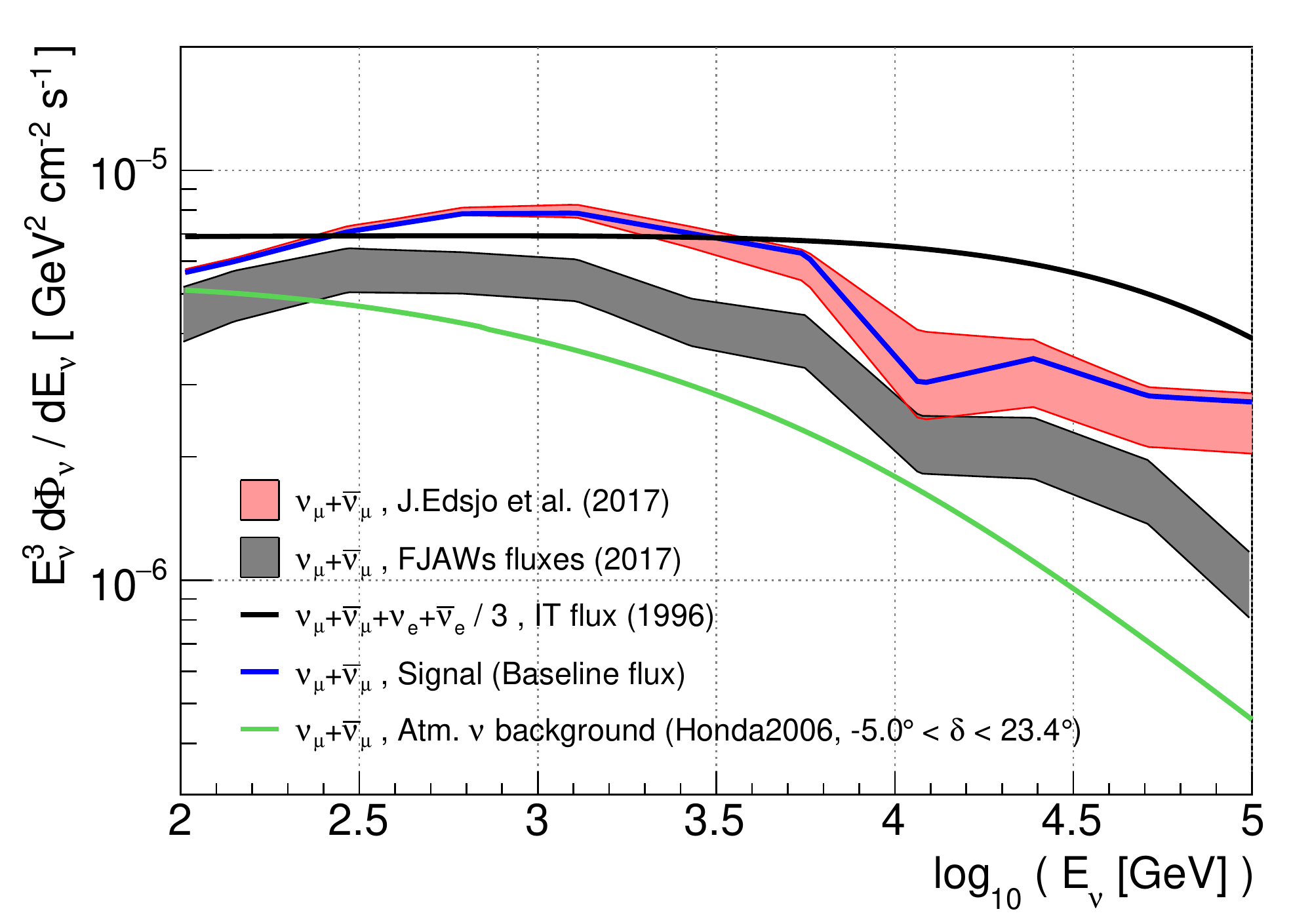}
\hfill
\caption{\label{fig:FluxComp}Predicted energy spectra of $\nu_{\mu}+\bar{\nu}_{\mu}$ at Earth. The energy spectra are integrated over the solid angle of the solar disk. The fluxes of SA$\nu$s are averaged along energy bins to smear out the effects of neutrino oscillation. The blue line is the baseline energy spectrum for systematic studies. The shaded areas cover the range of predictions from each reference (red for~\cite{Edsjo:2017kjk} and gray for~\cite{FJAWs2017}). The black line is the result of~\cite{IT1996} divided by a factor of three for neutrino oscillations. The green line is the Honda 2006 flux prediction~\cite{Honda2006} for terrestrial atmospheric neutrinos, which is time-averaged for the period when the Sun is below the horizon. It is added to demonstrate that the SA$\nu$ spectrum could be harder than that of neutrinos from the cosmic-ray interactions in the Earth's atmosphere.}
\end{figure}

In Fig.~\ref{fig:FluxComp}, the $\nu_\mu+\bar{\nu}_\mu$ neutrino flux predictions as well as their uncertainties are shown as the shaded regions. The range of the shaded gray area spans the energy spectra of the results published in~\cite{FJAWs2017,FJAWs_Public}. The red region represents the simulation results obtained by running the built-in codes in \textsf{WIMPSim}~\cite{Edsjo:2017kjk,WIMPSim}. 
Neutrino oscillations are fully taken into account when propagating the neutrinos from the Sun to the Earth and appear as wiggles in the theoretical flux predictions.
The oscillations of these high-energy neutrinos, however, cannot be resolved due to the limited energy resolution of IceCube. 
In Fig.~\ref{fig:FluxComp} we compare the energy spectra from a set of representative models~\cite{Edsjo:2017kjk, FJAWs2017}. For the visualization, we sample the flux predictions with coarse energy bins, that were chosen to reflect IceCube's energy response function.
Only the parametrized energy flux from Ref.~\cite{IT1996} (IT1996) did not include neutrino oscillations. Ref.~\cite{Fogli2006} has shown that if the primary flavor ratio of SA$\nu$s $(\nu_e:\nu_\mu:\nu_\tau)$ is $(1:2:0)$, it would be roughly close to $(1:1:1)$ at Earth. The flavor ratio of the IT1996 fluxes integrated in the range $(10^{2.0},10^{7.0})$ GeV is $(0.92,2.08,0)$. For simplicity, IT1996 fluxes for $\nu_\mu+\bar{\nu}_\mu+\nu_e+\bar{\nu}_e$ are divided by a factor of 3 to apply neutrino oscillation effect, shown in Fig.~\ref{fig:FluxComp} as the black line. Newer reference fluxes~\cite{Edsjo:2017kjk,FJAWs2017} already include the effect of the oscillations.

We measure the flux normalization of the SA$\nu$s in this analysis. A comparison of signal predictions~\cite{Edsjo:2017kjk,FJAWs2017,IT1996} shows that the SA$\nu$ spectral shapes are similar enough that we are not expected to be sensitive to individual models. We therefore choose one representative baseline energy spectrum (shown as the blue line in Fig.~\ref{fig:FluxComp}). The baseline energy spectrum is chosen from~\cite{Edsjo:2017kjk} and uses the Hillas-Gaisser 3-generation model~\cite{H3A} for the primary cosmic-ray spectrum, a combination of the Serenelli~\cite{Serenelli} and the Stein \emph{et al.}~\cite{Stein} models for the solar density profile, and the normal mass ordering. 

Finally, we note that the current leading models neglect solar magnetic field effects. These effects influence cosmic-ray propagation and the cascade development, which in turn influence the neutrino signal. The effect of magnetic fields on cosmic-ray propagation can be indirectly measured through the absorption of cosmic rays in the Sun, which in turn makes a corresponding deficit of cosmic rays in the direction of the Sun. The so-called cosmic-ray Sun shadow has been observed by the Tibet air shower array, including a variation of the intensity correlated with the solar cycle~\cite{Amenomori:2013own}. IceCube also observed the Sun shadow and found a correlation with the sunspot number with a likelihood of 96\%~\cite{ICShadow2019}. The Sun shadow is sensitive to magnetic field models~\cite{1969SoPh,Hakamada1995,1986ApJ,Zhao1995} and recent works with numerically computed trajectories of charged cosmic rays confirm the observationally established correlation between the magnitude of the shadowing effect and both the mean sunspot number and the polarity of the magnetic field during a solar cycle~\cite{BeckerTjus:2019rqu}. In general, however, high-energy cosmic rays are expected to be energetic enough not to be influenced by magnetic fields. Therefore, only for neutrino production below 200~GeV~\cite{Seckel1991} or 1~TeV~\cite{Masip:2017gvw} is it expected to become significant. Theoretical works using HAWC's Sun shadow observation predict a factor of about two difference in SA$\nu$ flux between solar minimum and maximum at 200~GeV~\cite{Masip:2017gvw}.

\subsection{Background predictions and competing signals}
\label{Subsec:BkgExp}
Most events in IceCube are downward-going atmospheric muons from cosmic-ray air showers in the Earth atmosphere. These muons can be efficiently rejected by selecting events reconstructed upward, \textit{i.e.} with declination $\delta > -5 \degree$. The well-established event selection for the upward-going neutrino events has achieved a purity of 99.7\%~\cite{ICRC2017_Sample}. In the remaining sample, the main background arises from terrestrial atmospheric neutrinos produced by decays of mesons within cosmic-rays air showers. Another irreducible, but sub-dominant background is due to isotropic astrophysical neutrinos. They can be described by an unbroken power-law with a spectral index of $2.19\pm0.1$ and a flux normalization, $\Phi_{\mathrm{100~TeV}}=1.01^{+0.26}_{-0.23}\cdot10^{-18}\mathrm{GeV^{-1}cm^{-2}s^{-1}sr^{-1}}$ at 100~TeV, obtained from fits to the data~\cite{ICRC2017_Sample}. The astrophysical neutrinos are included as a background in this analysis, but the uncertainties of the best-fit parameter values are negligible due to it's small contribution to the background rate.
 
Neutrinos from dark matter annihilations in the Sun could result in a competing signal that has been extensively searched for at neutrino telescopes~\cite{Danninger:2014xza,IC79SolarWIMP,IC86SolarWIMP,Choi:2015ara,Adrian-Martinez:2016gti,ICRC2017_SolarDM}. The expected neutrino spectra from solar dark matter strongly depend on the dark matter mass and annihilation channels. As dark matter annihilations are expected to occur in the center of the Sun, neutrino absorption becomes important for energies above 100~GeV and fluxes are  significantly attenuated above that energy. As a result, spectra are expected to be significantly different from that of SA$\nu$s~\cite{Kenny2017}. Purely based on event rate expectations at neutrino detectors, one can compute a sensitivity floor for indirect dark matter searches from the Sun~\cite{Kenny2017,FJAWs2017,Edsjo:2017kjk,Masip:2017gvw}. Past dark matter searches were not sensitive enough to have significant backgrounds from solar atmospheric neutrinos. However, in the near future they are expected to reach the neutrino floor from SA$\nu$s.

Another competing signal may arise from the interactions of cosmic rays with thermal solar photons. These can interact to form $\Delta^{+}$ baryons which quickly decay, producing muons and neutrinos from subsequent pion decays~\cite{Spencer2011}. The expected flux from $\Delta^{+}$ is small and few events are expected in IceCube, so we assume no contributions from the process in this analysis. Larger active volumes, like those proposed for IceCube-Gen2~\cite{Aartsen:2014njl,Aartsen:2020fgd}, may be needed to observe events from these interactions.

\begin{figure}[tbp]
\centering 
\includegraphics[width=.49\textwidth]{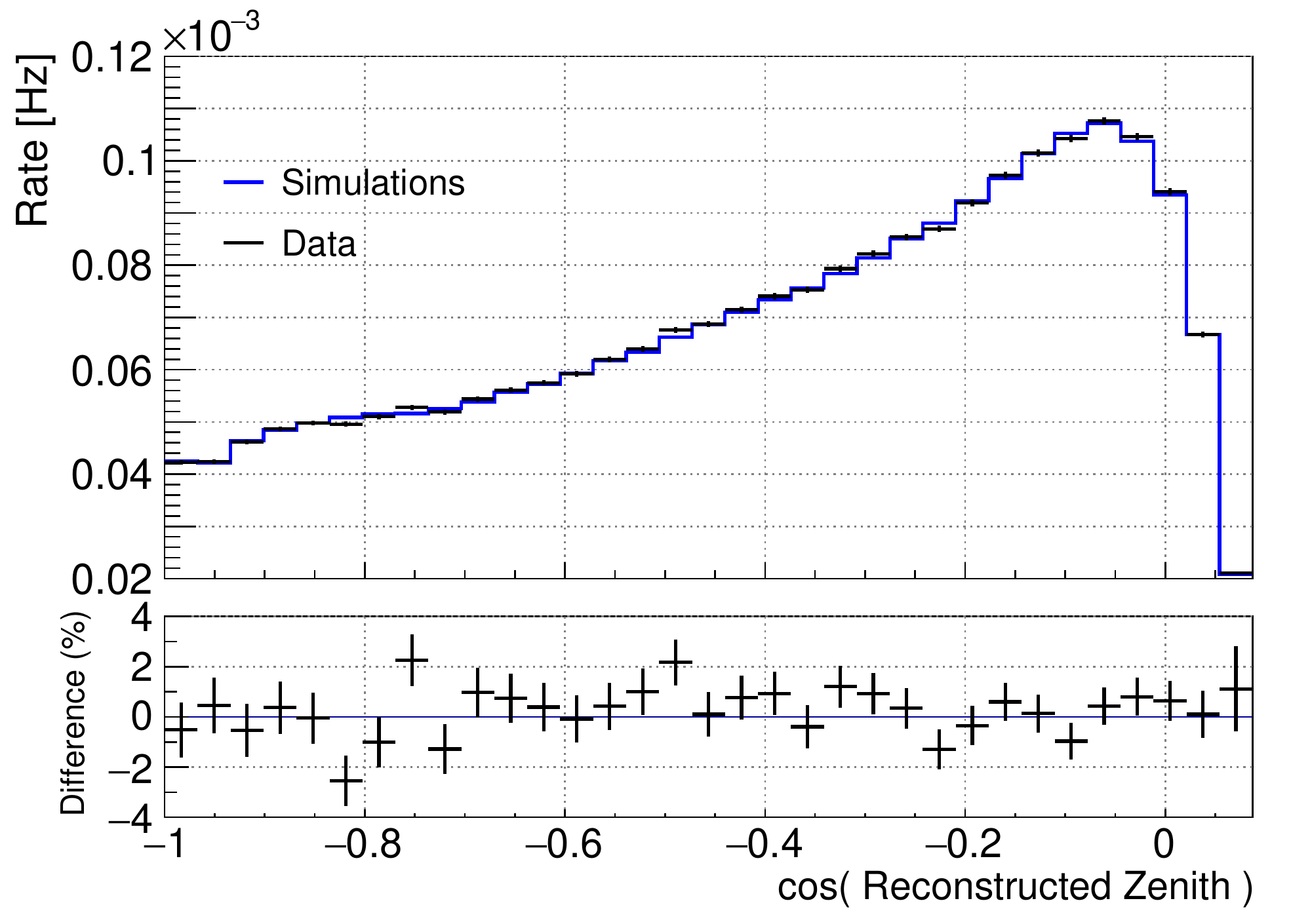}
\hfill
\includegraphics[width=.49\textwidth]{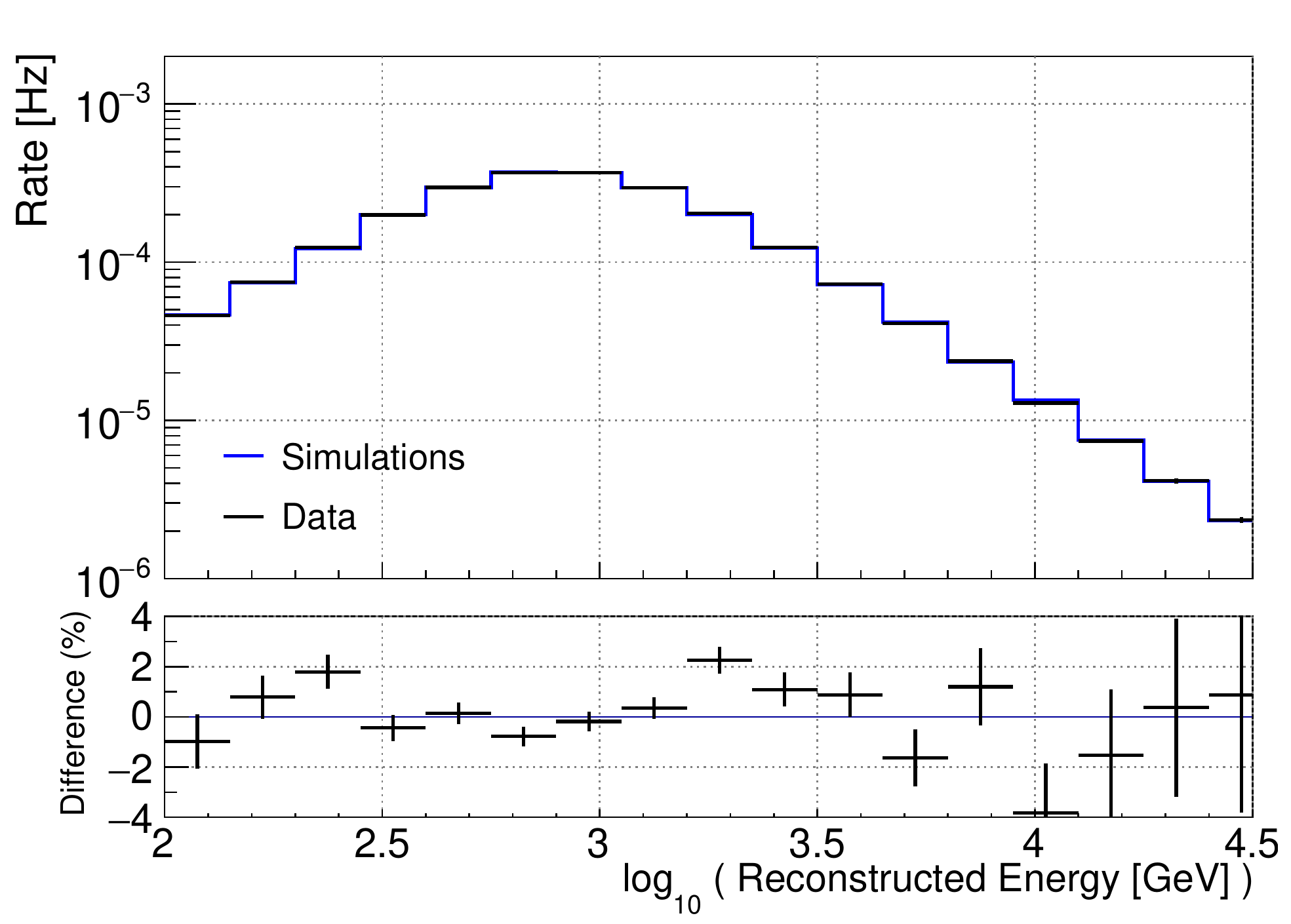}
\caption{\label{fig:DataMC}Reconstructed zenith angle and energy distributions for simulations (blue histogram) and data (black crosses, only statistical uncertainties shown). The difference is defined by (Data - Simulations) / Simulations as a percentage. The calculated rates are averaged over the analysis livetime.}
\end{figure}

\section{Data sample and simulations}
\label{sec:DataSim}
\subsection{Data sample}
\label{Subsec:Data}

A good angular resolution is necessary to search for SA$\nu$s because the angular size of the Sun is $\theta_\odot\sim0.27\degree$. Muons traversing the entire detector are reconstructed with good angular resolution as kilometer-long tracks, so-called ``through-going muons.'' We restrict ourselves to IceCube's neutrino sample of predominantly through-going muons~\cite{ICPS2019} providing $1.0\degree$ and $0.6\degree$ median angular resolutions at 1~\textrm{TeV} and 10~\textrm{TeV} neutrino energy, respectively. As the events are not fully contained in the detector volume, the energy resolutions are limited to $\Delta\log_{10}(E/\mathrm{1GeV})\sim0.3$ and $\sim0.5$ at 1~\textrm{TeV} and 10~\textrm{TeV}, respectively.

The data samples consist of three sub-samples covering a total of seven years. There are three time periods: IC79-2010, IC86-2011, and IC86-(2012-2016). An optimized event selection has been used for each configuration. The ranges of the reconstructed energies are ($10^{2.2}$, $10^{7.2}$) GeV and ($10^{2.0}$, $10^{7.0}$) GeV for IC79-2010 and IC86-(2011,2012-2016). Events below the horizon (declination, $\delta > -5 \degree$) are selected to exclude atmospheric muon events. Unlike Ref.~\cite{ICPS2019}, we only consider events where the Sun is below the horizon, resulting in a total analysis livetime of 1406.62~days. 

Angular separation ($\theta_\odot$) is defined as an angular distance between the reconstructed position and the center of the Sun at the event trigger time. We use an IceCube internal software implementation that relies on the Positional Astronomy Library~(PAL)~\cite{2013ASPC..475..307J} to obtain the position of the Sun based on the Modified Julian Date and the IceCube detector location. The tracking of the Sun was cross checked with data from NOAA to verify that it agrees within 0.01 degrees. We define a Region of Interest (RoI) as a $5\degree$ window around the center of the Sun. The size of the RoI was optimized taking into account the signal sensitivity and the computational time.
The sensitivity only marginally improves for larger windows as the RoI contains 96\% of the reconstructed signal events.

\begin{figure}[tbp]
\centering 
\includegraphics[width=.49\textwidth]{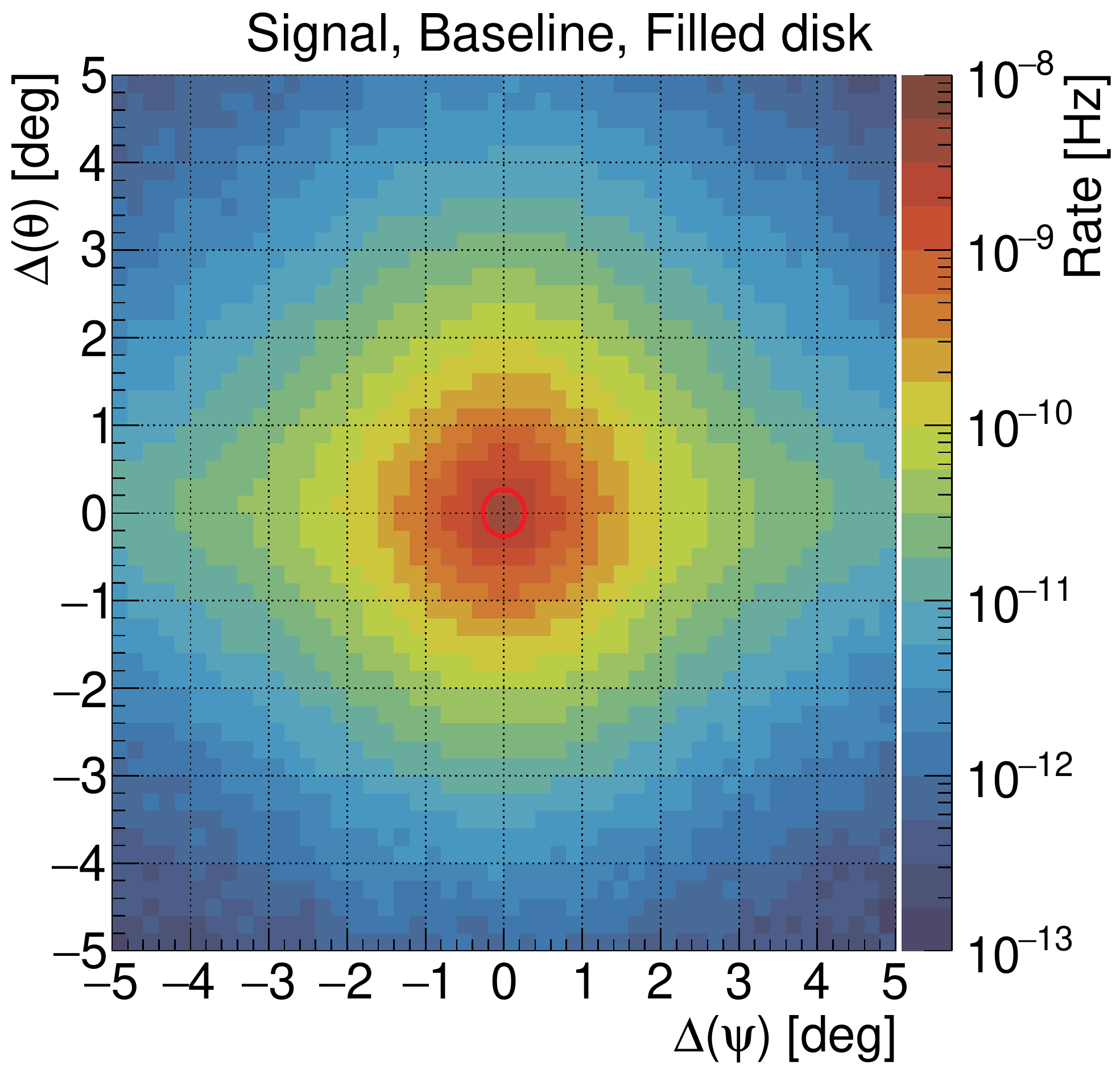}
\hfill
\includegraphics[width=.49\textwidth]{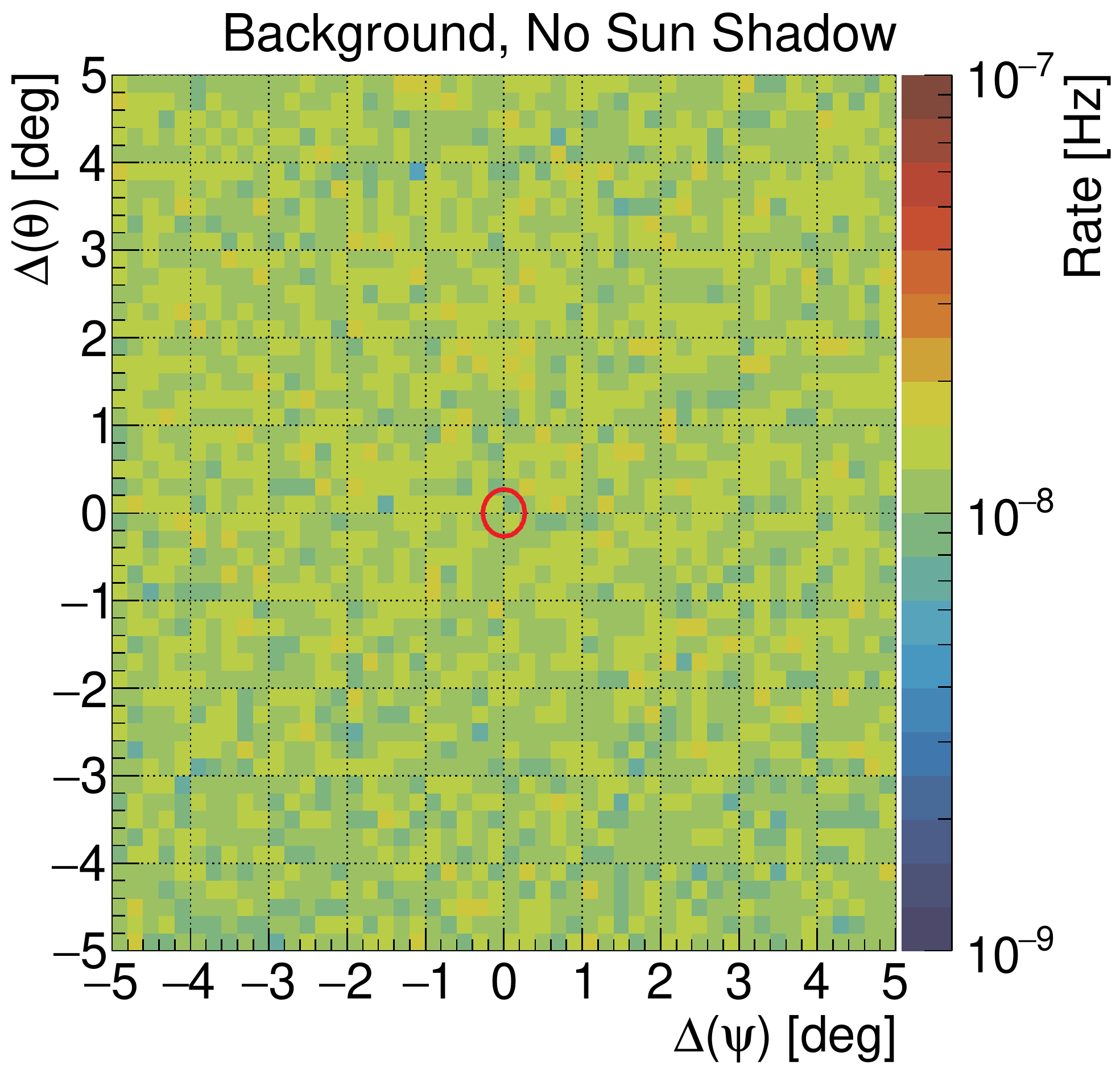}
\caption{\label{fig:2DAngDist}Reconstructed angular distributions from the baseline simulation assuming a \emph{filled disk} (see Sec.~\ref{Subsubsec:SourceDist}). The left and right plots show the expected signal and background, respectively. Axes are differences between the center of the Sun ($\Psi_{\textrm{Sun}}$, $\theta_{\textrm{Sun}}$) and the reconstructed directions ($\Psi_{\nu}$, $\theta_{\nu}$). The x-axis is the difference in azimuthal angle, $\Psi$, and the y-axis is the difference in zenith angle, $\theta$, while the z-axis is the event rate averaged for the total analysis livetime. The red circle represents the angular extent of the Sun. Note that the coordinate system does not directly project to the angular separation. The signal from the Sun is not fully symmetric in $\Theta$ and $\Psi$ due to the detector geometry and reconstruction uncertainties.}
\end{figure}

\subsection{Simulations}
\label{Subsec:Sim}
Simulations are used to obtain probability density functions~(PDFs) of the signal and background in the muon neutrino and muon anti-neutrino channels. Simulations samples are taken from the well established and tested IceCube point source analysis~\cite{ICPS2019}. We reuse these simulations but apply selection criteria on the angular separations within the RoI and reweight them with the effective analysis livetime. The background expectations are constructed using simulation, weighted to best-fit parameters for atmospheric and astrophysical neutrino backgrounds found from previous fits to data~\cite{ICRC2017_Sample}. In Fig.~\ref{fig:DataMC}, the comparisons between the simulations for terrestrial atmospheric neutrinos and the total data samples are shown for the reconstructed zenith angle and energy distributions. The simulation samples and the data samples are well-matched within 4\% differences.

Signal simulations are obtained by re-weighting the simulated events with the given SA$\nu$ energy spectrum for muon neutrinos (see Sec.~\ref{Subsec:SigExp}). The angular separations between the center of the Sun and the events are calculated. The azimuthal directions of the signal events are uniformly scrambled. Events are also randomized in zenith using the probability distribution as a function of angular distance from the center of the Sun for the given source hypothesis. We account for the movement of the Sun in zenith by weighting events using the fraction of livetime spent by the Sun in 30 zenith bins from $85\degree$ to $113.4\degree$. In Fig.~\ref{fig:2DAngDist}, two-dimensional angular distributions are shown for the baseline signal and background assumptions.

\section{Analysis}
\label{sec:Analysis}
\subsection{Unbinned likelihood analysis}
\begin{figure}[tbp]
\centering 
\includegraphics[width=.49\textwidth]{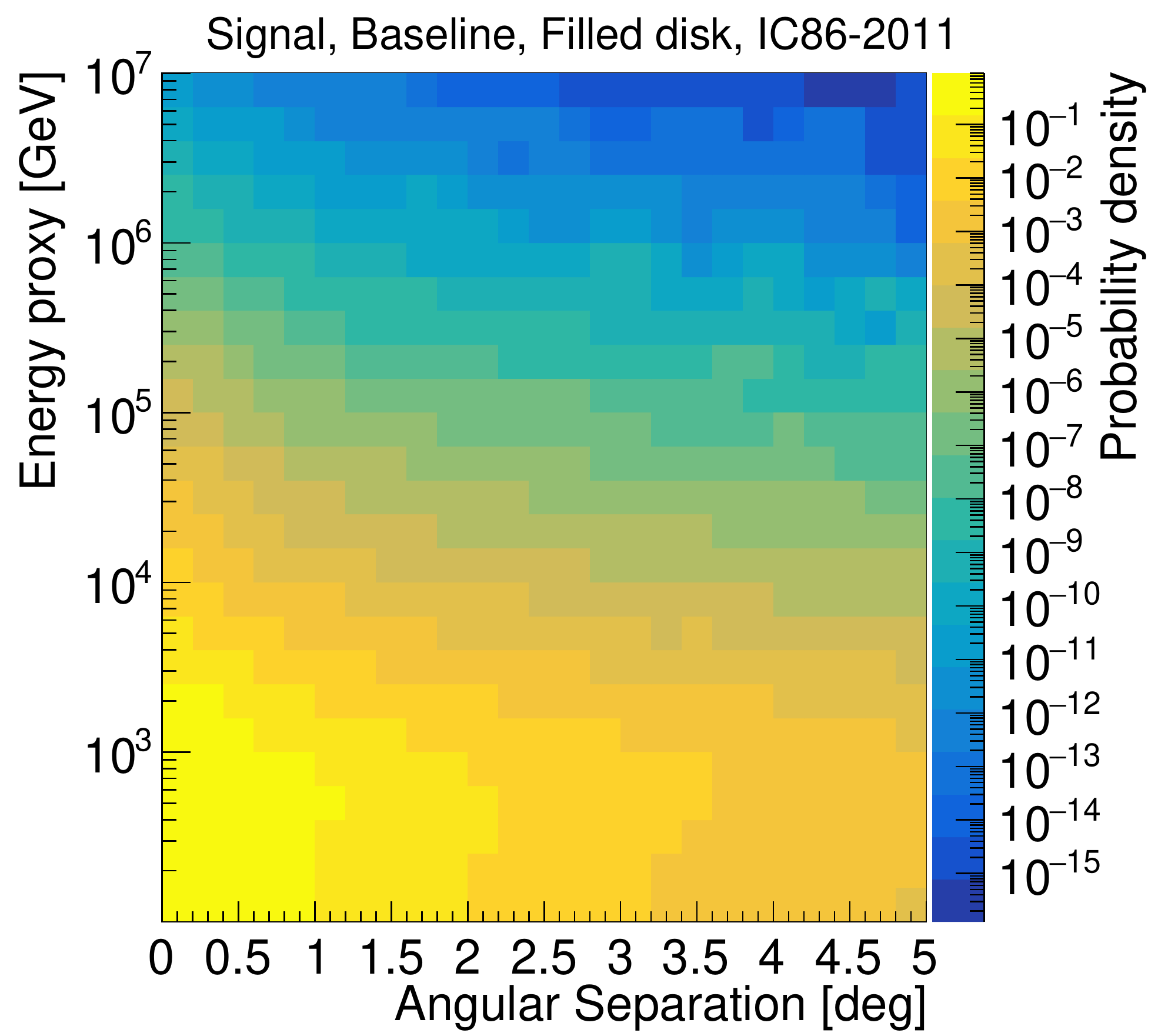}
\hfill
\includegraphics[width=.49\textwidth]{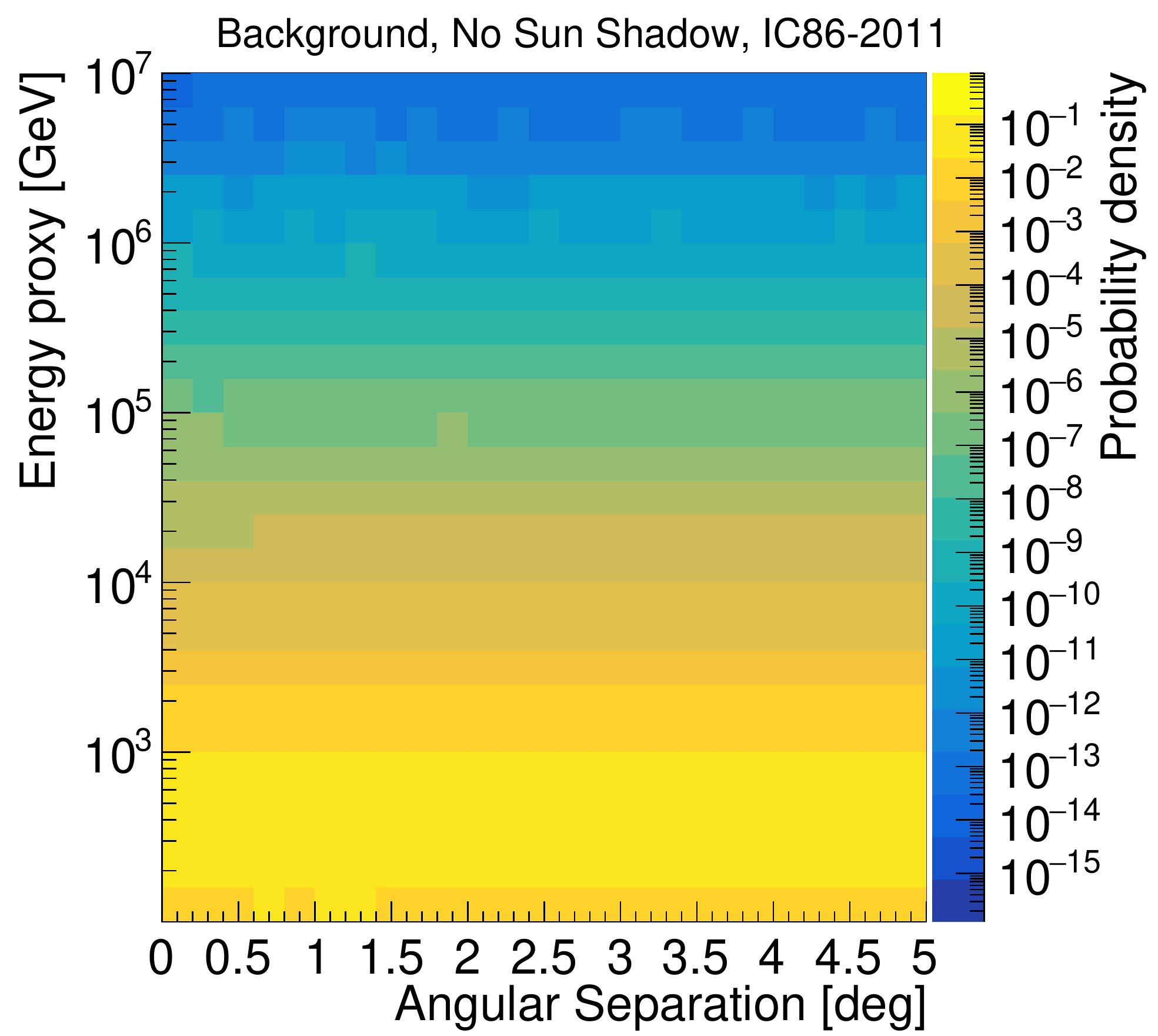}
\caption{\label{fig:ProbDist}Examples of PDFs in the likelihood functions. The left plot is a signal PDF, while the right plot shows the background PDF for IC86-2011. The probability densities are normalized to 1 in the RoI of the angular separation and energy proxy range. The energy proxy range is all ranges of the reconstructed muon energies in the simulation samples. The binning of the energy proxy is in 0.2 in the log of the reconstructed energy.}
\end{figure}

An unbinned likelihood method~\cite{AnaMeth2008} is applied to find evidence of SA$\nu$s in seven years of the data sample. The likelihood function, $L_{j}$, for each sub-sample, j, is defined by
\begin{equation}
\label{eq:LLHConf}
L_{j}(n_{s,\,j};M_{sig})=\prod_{i}^{n_{\textrm{tot},\,j}}\{\frac{n_{\textrm{s},\,j}}{n_{\textrm{tot},\,j}} \cdot p_{\textrm{sig},\,j}(\theta_i,E_i;M_{\textrm{sig}})+(1-\frac{n_{s,\,j}}{n_{\textrm{tot},\,j}}) \cdot p_{\textrm{bkg},\,j}(\theta_i,E_i)\},
\end{equation}
where $j$ is the index of the sub-sample, $i$ is the event index, $n_{\textrm{tot},\,j}$ is the total number of events and $n_{\textrm{s},\,j}$ is a number of signal events. For each event, $\theta_i$ is the angular separation to the Sun and $E_i$ is the reconstructed muon energy~\cite{Abbasi:2012wht}. The function of $p_{\textrm{sig},\,j}$ and $p_{\textrm{bkg},\,j}$ are the signal and background PDFs evaluated at the location of each event, respectively. In Fig.~\ref{fig:ProbDist}, PDFs of the IC86-(2012-2016) sub-sample are shown. The PDFs are obtained from the simulations and the corresponding likelihood functions are used to study a particular energy spectrum $M_{\textrm{sig}}$. We combine different sub-samples with a uniform signal emission and use the maximum likelihood estimator to estimate the signal strength. The total likelihood function, $L$, is a multiplication of the likelihood functions, $L_{j}$, for the three sub-samples mentioned in Sec.~\ref{Subsec:Data}. The fractions ($f_j$) of the total expected signal events for each sub-sample are calculated: $f_j = \bar{n}_{\textrm{s},\,j} / \Sigma_{k}\bar{n}_{\textrm{s},\,k}$ where $\bar{n}_{s,\,j}$ is an expected number of signal events from the simulations. The total likelihood function is redefined as a function of the total signal strength $\mu$ with converting $n_{\textrm{s},\,j}$ to $\mu f_j$:
\begin{equation}
\label{eq:LLHTotal}
L(\mu)=\prod_{j}\prod_{i}^{n_{\textrm{tot},\,j}}\{\frac{\mu f_j}{n_{\textrm{tot},\,j}} \cdot p_{\textrm{sig},\,j}(\theta_i,E_i|M_{\textrm{sig}})+(1-\frac{\mu f_j}{n_{\textrm{tot},\,j}}) \cdot p_{\textrm{bkg},\,j}(\theta_i,E_i)\}.
\end{equation}

The theoretical distribution of flux across the solar disk is expected to depend on neutrino energy via the energy dependence of IceCube reconstructions. To include this correlation, two dimensional PDFs, shown in Fig.~\ref{fig:ProbDist}, are used to model the signal and background distributions in the likelihood functions.

We define the test statistic (\textit{TS}) 
\begin{equation}
\label{eq:TS}
\begin{split}
TS&=\,\,\,\:2\ln{L(\hat{\mu})/L(0)} \quad \textrm{for} \quad\hat{\mu}\ge0 \\
&=-2\ln{L(\hat{\mu})/L(0)} \quad \textrm{for} \quad\hat{\mu}<0,
\end{split}
\end{equation}
as the likelihood ratio between the best-fit value and the null hypothesis. The range of $\hat{\mu}$ is not restricted to positive values. Therefore, we can track the sign of $\hat{\mu}$ to separately determine sensitivities for a positive or negative signal strength. The negative signs of $\hat{\mu}$ can appear when the alternate hypothesis represents an under-fluctuation relative to the background prediction, especially that the under-fluctuation can be enhanced by the Sun shadow, see Sec.~\ref{Subsubsec:Sunshadow}.

\subsection{Sensitivity calculations}
\label{Subsec:Sensitivity}
Pseudo-experiments are conducted to obtain the \textit{TS} distribution for a given hypothesis. Each pseudo-experiment consists of mock samples generated by random sampling based on each PDF of a certain hypothesis. The number of signal and background events are random variables that are Poisson distributed. The mean of the Poisson distribution for the number of background events is given by the expected number of events from the simulations, $\bar{n}_{\textrm{bkg}}=1147.4$ in the RoI. Depending on the hypotheses, the mean for the signal $\Bar{\mu}$ is scaled, \textit{e.g.} $\Bar{\mu}$=0 for the null hypothesis and $\Bar{\mu}=C_s \cdot \bar{n}_{\textrm{sig}}$, where $C_s$ is a scale factor to test $C_s$ times larger signal hypotheses. $\bar{n}_{\textrm{sig}}$ is the expected number of signal events determined by combining a given signal model with the simulated detector response. The expected number of background events, $\bar{n}_{\textrm{bkg}}$, and signal events, $\bar{n}_{\textrm{sig}}$, also change with the PDFs according to the hypotheses chosen.

\begin{figure}[tbp]
\centering 
\includegraphics[width=.7\textwidth]{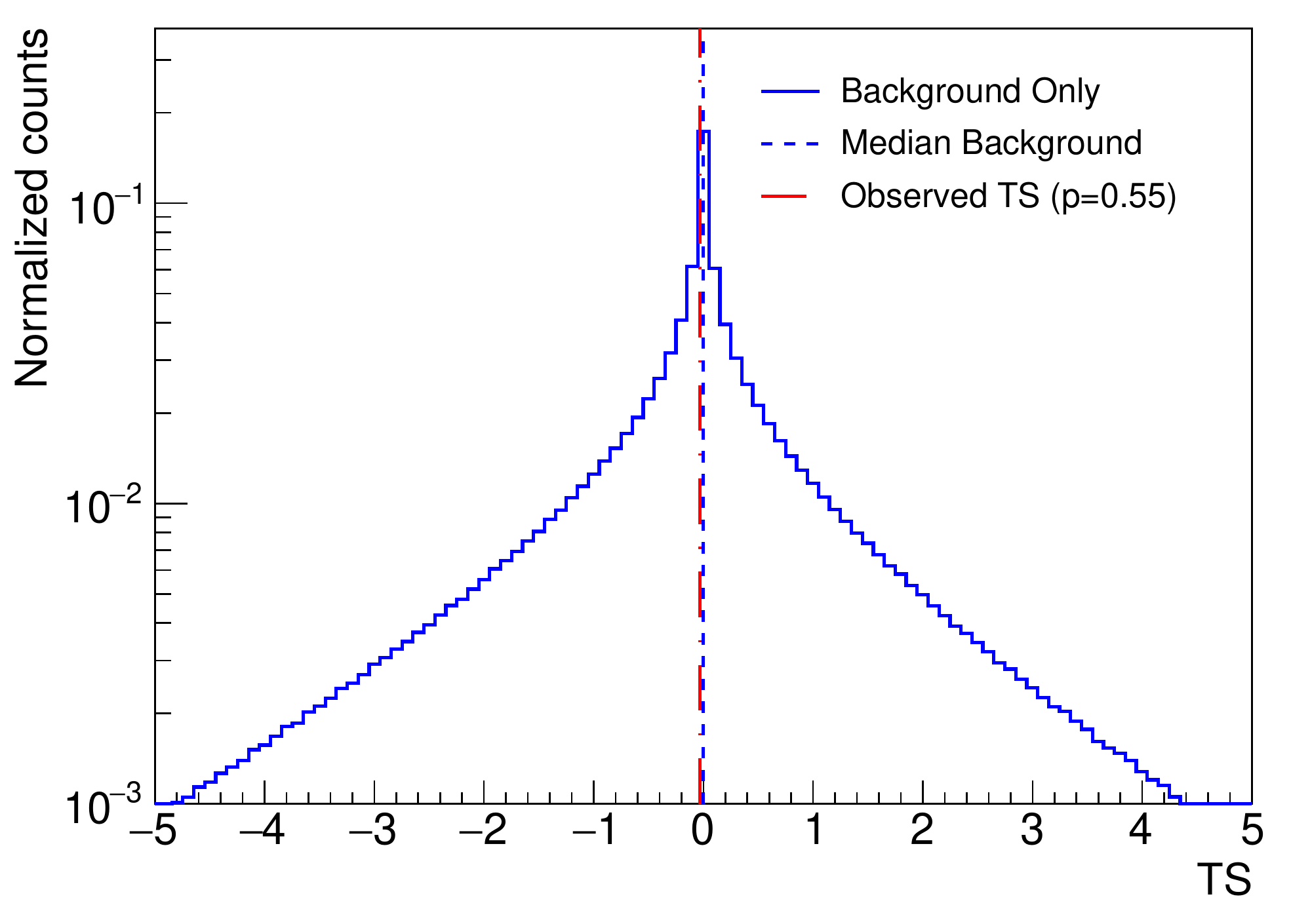}
\hfill
\caption{\label{fig:TSDist}The blue histogram is the \textit{TS} distribution for the background only hypothesis, normalized to 1. The \textit{TS} is negative when the mock sample of the pseudo-experiment contains an under-fluctuation relative to the expected background rate. The blue dashed line is the median of the histogram. The red dashed line is the observed \textit{TS} value from the experimental data.}
\end{figure}

In Fig.~\ref{fig:TSDist}, the blue histogram is the \textit{TS} distribution  for the background-only hypothesis. Negative TS values appear when the likelihood function is maximized with a negative signal strength, due to under-fluctuations in the background rate. The median of the histogram, indicated by the vertical dashed blue line in Fig.~\ref{fig:TSDist}, is close to zero. The 90\% confidence interval (C.I.) is obtained with the Feldman-Cousins method~\cite{FCInterval} for each alternate hypothesis. $\mu_{90}$ is defined by $\bar{\mu}$ of the Poisson mean when the minimum of the 90\% C.I. is larger than the median of the \textit{TS} distribution for the null hypothesis. The 90\% confidence level (C.L.) upper sensitivities are set with $\mu_{90}=C_{s,90} \cdot \bar{n}_{\textrm{sig}}$ for each SA$\nu$ flux model given by Refs.~\cite{Edsjo:2017kjk,FJAWs2017,IT1996}. PDFs are used in the likelihood functions and the random sampling for the mock samples of the pseudo-experiments. The sensitivities to each flux model are calculated with the corresponding PDFs.  The sensitivity to the baseline signal spectrum  is shown as part of our final result, see the red solid line in Fig.~\ref{fig:Limit}.
It is 12.8 times larger than the theoretical expectation from the baseline model flux~\cite{Edsjo:2017kjk}.

\subsection{Systematic uncertainties}
\label{Subsec:Systematic}
We investigate how the sensitivity of our analysis depends on different choices for flux distributions on the solar disk, oscillation parameters, the effect of the Sun shadow on the backgrounds and detector uncertainties. The differences between the sensitivities are quantified relative to the baseline model as systematic uncertainties.

\subsubsection{Flux distribution on the solar disk}
\label{Subsubsec:SourceDist}

\begin{figure}[t]
\centering 
\includegraphics[width=\textwidth]{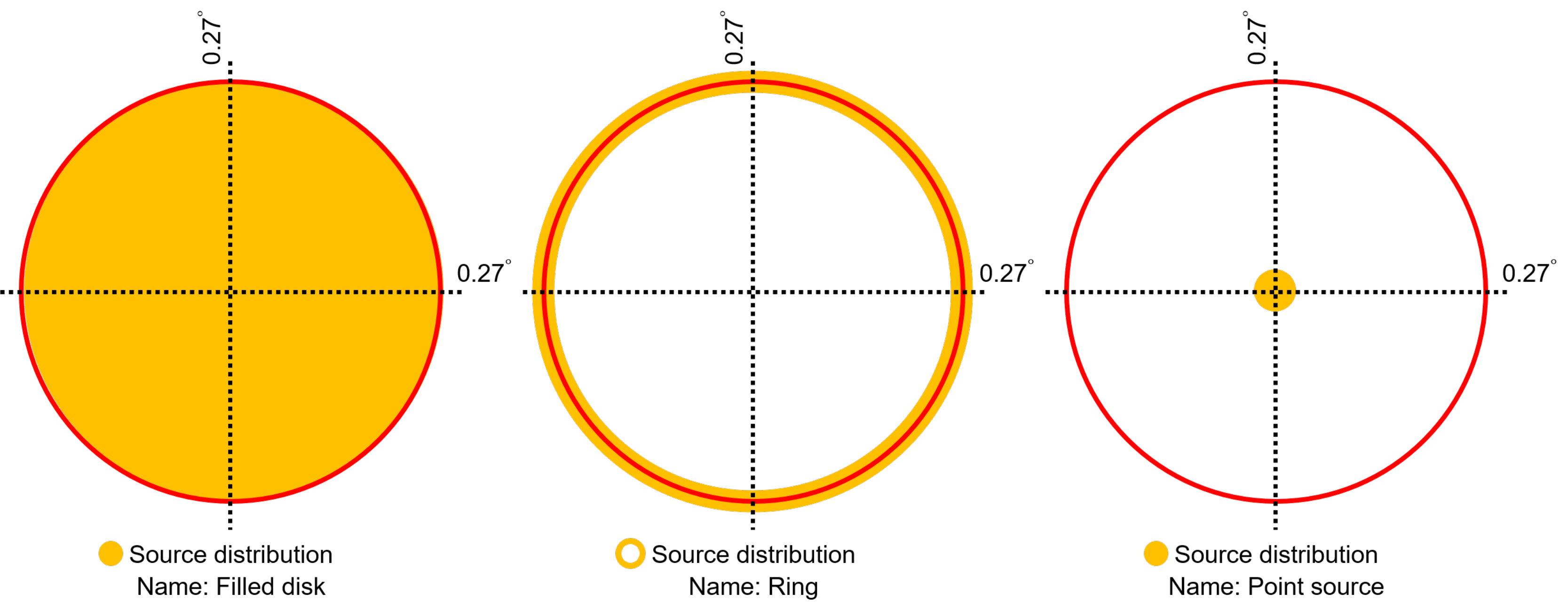}
\hfill
\caption{\label{fig:SourDist}Schematic diagrams for the extreme cases of the distribution of the neutrino flux on the solar disk. The red ring represents the scale of the Sun as $\theta_\odot=0.266 \degree$. The orange regions represent the distributions of the signal. Left: the signal is uniformly distributed in the solar disk (\emph{Filled disk}, used as baseline). Middle: the signal is emitted at the edge of the Sun (\emph{Ring}). Right: the signal only emanates from the center of the Sun (\emph{Point source}).}
\end{figure}

High-energy neutrinos above 1~TeV will be strongly suppressed when they propagate through the center of the Sun ($\theta_{\odot}~=~0\degree$), while the attenuation is much weaker at the edge of the Sun ($\theta_{\odot}~\simeq0.27\degree$). For instance, the survival probability is larger than $\sim90\%$ for neutrino energies below 100~TeV~\cite{Edsjo:2017kjk} at the edge. On the other hand, $\sim20\%~(\sim35\%)$ of 100~\textrm{GeV} neutrinos (anti-neutrinos) survive when they traverse the entire Sun at the center and they are almost completely absorbed above 1~TeV. Therefore, the high-energy events mostly arise from the edge of the Sun. 
The low-energy signal events emanate relatively uniformly over the solar disk while the high-energy neutrino signal is expected to have a dip in the central region. If we consider magnetic field effects, however, these could act differently due to cosmic-ray mirroring, providing additional contributions in particular toward the center of the disk. This would presumably lead to a more uniform distribution for low energies. As a model-independent method, we consider three extreme cases for the spatial distribution on the disk, shown in Fig.~\ref{fig:SourDist}. Our baseline model is \emph{Filled Disk} where the signals are uniformly distributed on the solar disk. The simplest assumption is that all neutrinos are coming from the center of the Sun, named \emph{Point Source}. This leads to the best sensitivity with 3\% improvement compared to the baseline model. In contrast, \emph{Ring} assumes that the signals are only located at the edge of the Sun. For high-energy neutrinos, the distribution is expected close to \emph{Ring} due to absorption across the solar core. The fluxes are equally normalized for all cases. The true spatial distribution depends on the neutrino energy. We adopt the \emph{Filled Disk} as our baseline and use the two other extreme cases to evaluate the systematic uncertainty due to the choice of the source distribution.

\subsubsection{Neutrino oscillation parameters}
\label{Subsubsec:nuOsc}
After SA$\nu$s are produced in the Sun, the neutrinos oscillate while propagating to the Earth. The uncertainties on the oscillation parameters can alter the energy spectrum. The oscillation parameters used for the baseline energy spectrum are listed in the column denoted as baseline in Tab.~\ref{tab:1}. We checked the effect of varying the parameters by 1$\sigma$ on the energy spectrum. Also, the best-fit values for $\theta_{23}$ in the both octants are considered. The uncertainties on neutrino oscillation parameters are treated as systematic uncertainties but the sensitivities for each mass ordering are calculated separately for the energy spectra given by Ref.~\cite{Edsjo:2017kjk}.

\renewcommand{\arraystretch}{1.3}
\begin{table}[t]
\centering
\begin{tabular}{|c|r|r|r|r|}
\hline
 & Baseline & \multicolumn{2}{|r|}{$1\sigma$} & Octant\\
\hline
$\Delta m^2_{32}/10^{-3}$ (eV\textsuperscript{2}) & $2.51$ & $\pm0.05$ & $-2.56\pm0.04$ & \\
\hline
$\Delta m^2_{21}/10^{-5}$ (eV\textsuperscript{2}) & $7.53$ & $\pm0.18$ & & \\
\hline
$\sin^{2}{\theta_{12}}/10^{-1}$ & $3.07$ & $\pm0.13$ & & \\
\hline
$\sin^{2}{\theta_{13}}/10^{-2}$ & $2.12$ & $\pm0.08$ & & \\
\hline
\multirow{4}{*}{$\sin^{2}{\theta_{23}}/10^{-1}$} & \multirow{2}{*}{$4.17$} & $+0.25$ & $4.21+0.33$ & \multirow{2}{*}{Octant 1} \\
& & $-0.28$ & $-0.25$ & \\   
\cline{2-5}
& & $6.21$ & $5.92+0.23$ & \multirow{2}{*}{Octant 2} \\
& & $5.67$ & $-0.30$ & \\
\hline
Mass Ordering & \multicolumn{2}{|c|}{Normal} & \multicolumn{1}{|c|}{Inverted} & \\
\hline
\end{tabular}
\caption{\label{tab:1}The neutrino oscillation parameters for the flux calculation in \textsf{WIMPSim}. The values are the best-fit results of Ref.~\cite{PDG2018}. The second column named ``Baseline'' lists the parameters for the baseline energy spectrum. The energy spectra for the signal are obtained by \textsf{WIMPSim}, where we independently vary a parameter in 1 $\sigma$ region and allow $\sin^{2}{\theta_{23}}$ to lie in either octants.}
\end{table}

\subsubsection{Sun shadow effect on the backgrounds}
\label{Subsubsec:Sunshadow}
The cosmic-ray flux coming from the direction of the Sun is expected to be less than that from other directions because cosmic rays are absorbed by the Sun itself, creating what is referred to as the Sun shadow. The Sun shadow effect has been observed as a deficit of atmospheric muons~\cite{ICShadow2019,Aartsen:2020hzn}. The angular extent of the deficit can be approximated with one-sided Gaussian functions for each season. While we use the case without the Sun shadow as the baseline, the Sun shadow should also reduce the terrestrial atmospheric neutrinos which are the dominant background in this analysis. However, the deficit of the terrestrial atmospheric neutrinos by the Sun shadow has not been studied before. To take this into account, we assume that the neutrino rate decreases with the same fractional strength and angular dependence as the muons studied in Ref.~\cite{ICShadow2019}. In simulations, the terrestrial atmospheric neutrino events are re-weighted with the one-sided Gaussian functions of Eq.~\ref{eq:Shadow}:

\begin{equation}
\label{eq:Shadow}
\Delta N_{\nu} / N_{\nu} = -A\cdot \exp{(-\theta^{2}_{\odot}/2\sigma^2)} \,
\begin{cases}
\, A = 0.11 \,,\, \sigma=0.53\degree \text{ for IC79-2010} \\
\, A = 0.08 \,,\, \sigma=0.49\degree \text{ for IC86-2011} \\
\, A = 0.07 \,,\, \sigma=0.57\degree \text{ for IC86-(2012-2016)},
\end{cases}
\end{equation}
where $A$ and $\sigma$ are the best-fit parameters for the observed muon deficits by IceCube~\cite{ICShadow2019}. The parameters for IC86-(2012-2016) are averaged values to match time period of the sub-sample (see Sec.~\ref{sec:DataSim}). The parameters are time-dependent because they are correlated with solar activities~\cite{ICShadow2019,BeckerTjus:2019rqu}. Uncertainties on the best-fit parameters $A$ and $\sigma$ are $\sim10\%$. Although the deficit of the terrestrial atmospheric neutrinos is expected, we choose to set the baseline background predictions without the Sun shadow effect. Thereby, we have taken a conservative approach in the analysis, which requires a larger signal for a discovery. The Sun shadow effect is included as a systematic uncertainty.

\subsubsection{Uncertainty calculations}
\label{Subsubsec:UncertaintyCal}

\renewcommand{\arraystretch}{1.5}
\begin{table}[t]
\centering
\begin{tabular}{|l|r|l|}
\hline
Sources & \makecell[r]{Systematic \\ Uncertainties} & Comments \\
\hline
\hline
Detection efficiency of DOM & (-15\%, + 11\%) & \\
\hline
\makecell[l]{Absorption and \\ scattering efficiency of ice} & (-5\%, +12\%) & \\
\hline
Photo-nuclear interaction & (-3\%, +4\%) & Uncertainties for high-energy muons \\
\hline
Morphology & (-3\%, +3\%) & \textit{Filled disk} $\rightarrow$ \textit{Ring}, \textit{Point source} \\
\hline
Sun shadow & -11\% & w/o Sun shadow $\rightarrow$ w/ Sun shadow\\
\hline
$\nu_\tau$,$\bar{\nu}_\tau$ contribution & 4\% & $\nu_\mu$,$\bar{\nu}_\mu$ $\rightarrow$ $\nu_\mu$,$\bar{\nu}_\mu$,$\nu_\tau$,$\bar{\nu}_\tau$ \\
\hline
$\nu$ oscillation parameters & <1\% & \\
\hline
\hline
Total &  (-19.7\%, +17.8\%) & Assumes uncorrelated uncertainties\\
\hline
\end{tabular}
\caption{\label{tab:Sys}Summary of impact on sensitivity. A plus sign corresponds to an improved sensitivity.}
\end{table}

The uncertainties of the sensitivities for the source distributions, neutrino oscillation parameters in the signal prediction and the Sun shadow effect in the background prediction are calculated with the same simulation samples. We randomize the positions of each signal event from the distribution of locations allowed by each Sun model. The same simulations are used for the oscillation parameters and the Sun shadow effect, but the weights in the simulations are modified with the corresponding energy spectra and the deficit rates, respectively. Another main systematic uncertainty arises from standard detector uncertainties including the optical efficiency of DOMs for the Cherenkov light detection~\cite{ICPMT2010}, the optical absorption and scattering properties of the ice~\cite{Aartsen:2013vja}, and the uncertainties on photo-nuclear interaction cross sections of high-energy muons~\cite{InterA,InterB,InterC,InterD,InterE,InterF,InterG}. The same simulations and detector uncertainties are used as in Ref.~\cite{ICPS2019}. 

We calculate the sensitivities for the systematic uncertainties as alternate hypotheses. The signal and background PDFs for the baseline are tested against events sampled from PDFs generated from variations of the systematic uncertainties. The uncertainties of the sensitivities are quantified as the differences of the scale factor $C_{s,90}$ when the energy spectrum of the signal is identical to the baseline. The uncertainties of the neutrino oscillation parameters change the shape of the SA$\nu$ spectra. To quantify the systematic uncertainties, the differences of the $\mu_{90}$ are used for the uncertainties of the neutrino oscillation parameters.
    
Detector uncertainties give the largest systematic uncertainties in this analysis. When we vary the efficiency of DOMs by $\pm10\%$, the sensitivity changes in the range of $(-15,+11)\%$, with positive values indicating improved sensitivity. Simulation data sets with different optical absorption and scattering lengths of the ice are available for the values of $(+10,0)\%$, $(0,+10)\%$ and $(-7.1,-7.1)\%$. We used those simulations to estimate the uncertainties due to ice properties and they affect the sensitivity by $-5\%$ to $12\%$. The same simulation samples in Ref.~\cite{ICPS2019} are used for studying photo-nuclear interaction models of high energy muons~\cite{InterA,InterB,InterC,InterD,InterE,InterF,InterG}. This leads to uncertainties on the sensitivity ranging from $-3\%$ to $4\%$. We consider this estimate to be conservative as the models represent extreme cases which are outdated~\cite{ICPS2019}.

The neutrino oscillation parameters introduce less than $1\%$ uncertainty. Tests of the \textit{Ring} and \textit{Point source} emission distributions yield an uncertainty of $\pm3\%$.
As an uncertainty of the background predictions, the Sun shadow effect has been studied. Compared to the baseline background prediction, the number of background events decreases near the Sun due to the shadow effect. As a result, the likelihood function is maximized with a negative signal strength as the under-fluctuation of the null hypothesis. With the Sun shadow included in the background prediction, a larger $\bar{\mu}$ is necessary to obtain the same sensitivity level with the baseline prediction, hence it increases by 11\%.

The simulations assume only muon neutrino and muon anti-neutrino interactions. The fluxes of $\nu_\tau+\bar{\nu}_\tau$ for SA$\nu$s can be calculated with \textsf{WIMPSim} for the baseline energy spectrum. Similar amplitudes of $\nu_\tau+\bar{\nu}_\tau$ and $\nu_\mu+\bar{\nu}_\mu$ fluxes are expected through neutrino oscillations from the Sun to the Earth. However, the detection efficiency for $\nu_\tau+\bar{\nu}_\tau$ is much smaller. When we add the additional contribution on the signal by $\nu_\tau+\bar{\nu}_\tau$ using the simulations used in Ref.~\cite{ICPS2019}, the sensitivities improves by $4\%$. The contribution from $\nu_e+\bar{\nu}_e$ is negligible due to the event selection strongly favoring track-like events.

Assuming fully uncorrelated uncertainties, the total uncertainty on the median sensitivity is $-19.7\%$ to $+17.8\%$ and is dominated by detector uncertainties. Table~\ref{tab:Sys} summarizes the systematic studies. The systematic uncertainties on sensitivities are shown as the red region as part of the final result in Fig.~\ref{fig:Limit}. The systematic uncertainties are similar to those in a previous study with the same samples~\cite{ICPS2019} but the results are slightly distinct because we track the Sun rather than point sources at specific zenith angles.

\section{Results}
\label{sec:Results}

\begin{figure}[tbp]
\centering 
\includegraphics[width=.7\textwidth]{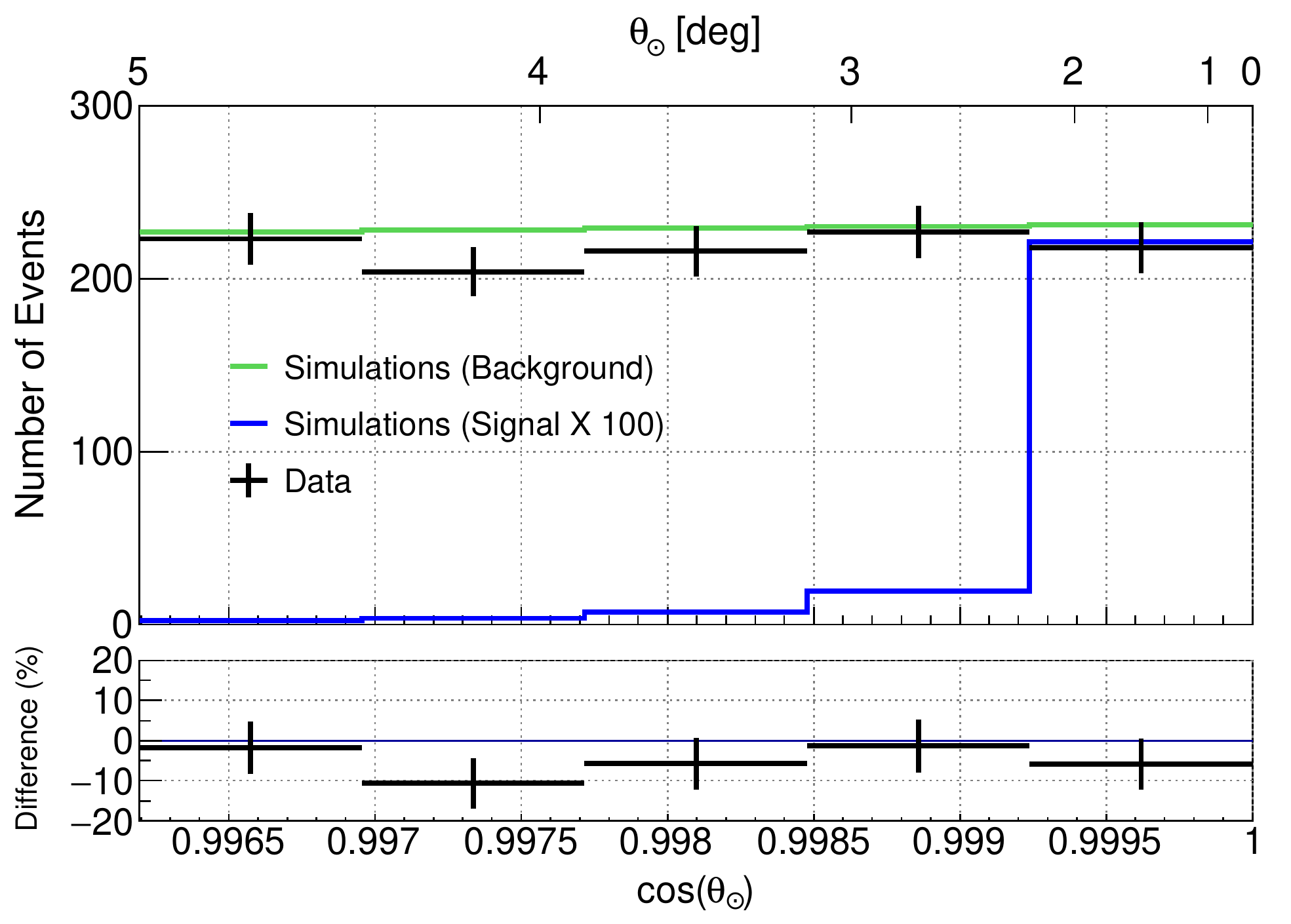}
\hfill
\caption{\label{fig:1DAngDist}Event distribution in the angular separation ($\theta_{\odot}$) within the RoI. Black crosses represent the experimental data, the green histogram shows the background prediction without the Sun shadow effect. The blue histogram shows the baseline signal prediction scaled by a factor of 100. The simulation and data are shown for the winter season. The difference is defined as (Data - Simulations) / Simulations. This figure is only used for visualization of the data. A binned representation of it is shown here, however, the analysis uses an unbinned likelihood method.}
\end{figure}

For visualization we show the observed data in the RoI and compare it to the background expectations in a binned representation, however, the analysis uses an unbinned likelihood method.
In the top panel of Fig.~\ref{fig:1DAngDist}, the angular distribution of the experimental data in the RoI is shown by the black crosses. The observed data are within 10\% of the background prediction (green histogram), and the number of events in the RoI is statistically compatible with the background expectation at $1.75\sigma$. The best-fit values $\hat{\mu}$ for all energy spectra are shown in Tab.~\ref{tab:Sensitivity} and are always negative, indicating an under-fluctuation in the data relative to the expected background. No evidence of SA$\nu$s is found in seven years of IceCube data. The observed \textit{TS} for the baseline signal prediction is the red dashed line in Fig.~\ref{fig:TSDist}. It is very close to the median of the \textit{TS} distribution for the null hypothesis, with an observed p-value of 0.55. Here, the p-value is defined as the area of the \textit{TS} distribution above the observed \textit{TS} value.

\begin{figure}[tbp]
\centering 
\includegraphics[width=1.0\textwidth]{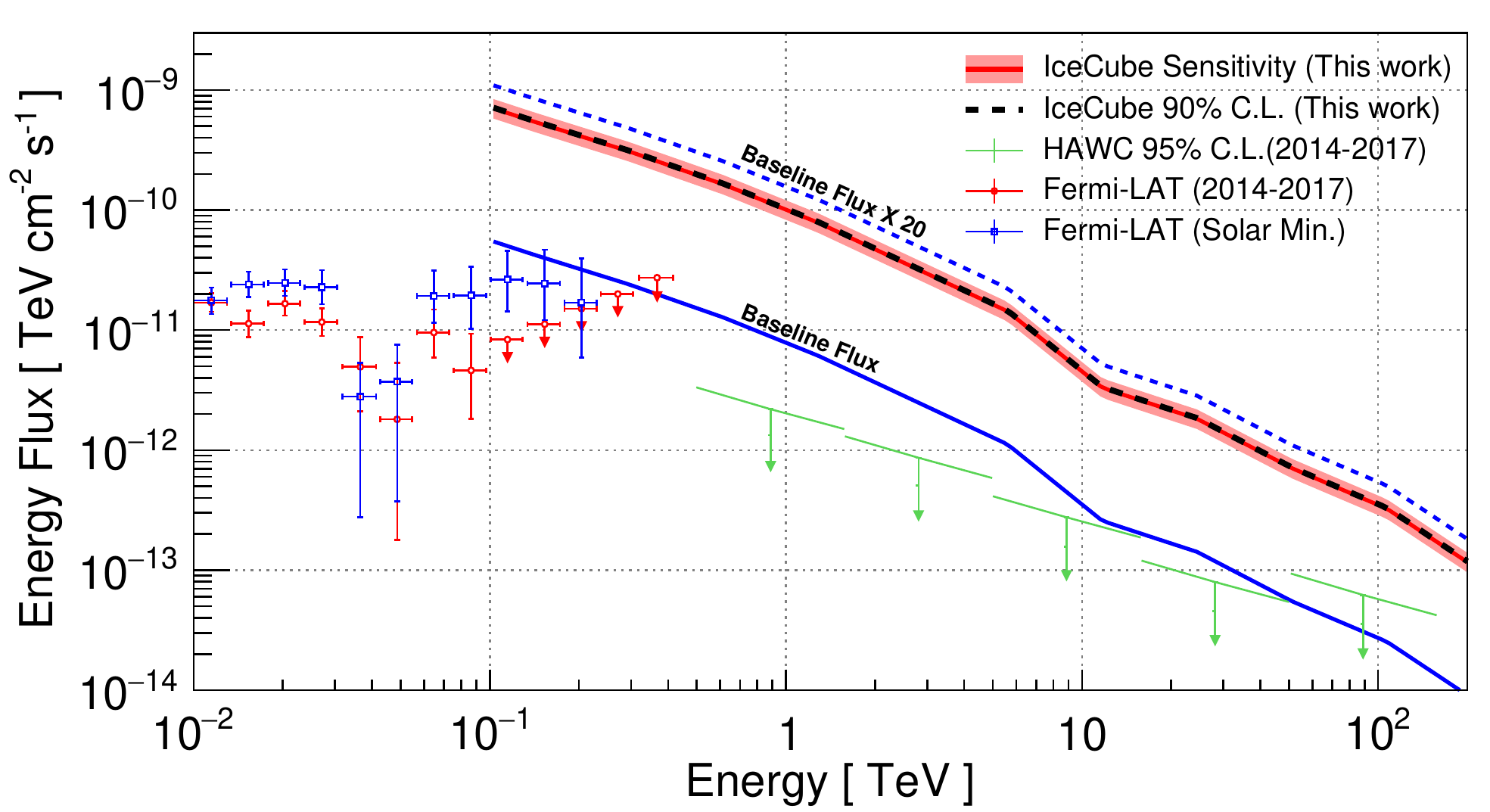}
\hfill
\caption{\label{fig:Limit}
IceCube 90\% C.L. upper limit is the black dashed line assumed the signal following the baseline flux expectation, the blue solid line. For comparison, the blue dotted line shows the baseline flux scaled by a factor of 20. The red shaded band illustrates the corresponding uncertainty of the baseline model. In addition, we include results from gamma-ray observations in the plot. Red and blue crosses are the observations of Fermi-LAT~\cite{Tim2018,Qing2018}; green points correspond to HAWC's 95\% C.L limit~\cite{HAWC2018,Zhou:2016ljf}.} 
\end{figure}

The observed p-value being larger than 0.5 indicates that there is a slight under-\-fluctuation in the background expectation. We place a 90\% C.L. upper limit for $\mu_{90}$, when the lower edge of the 90\% C.I. is larger than the observed \textit{TS} value. In Fig.~\ref{fig:Limit}, the black dashed line represents this limit. The values obtained for $\mu_{90}$ $(C_{s,90})$ are $36.5$ $(13.0)$. At 1~TeV, the limit on the flux normalization is $1.02^{+0.20}_{-0.18}\cdot10^{-13}$ $\mathrm{GeV^{-1}cm^{-2}s^{-1}}$ including the systematic uncertainties. Table~\ref{tab:Sensitivity} contains the full analysis results with limits on all SA$\nu$ flux models. The limits calculated on the basis of Ref.~\cite{Edsjo:2017kjk} and Ref.~\cite{FJAWs2017} turn out to rather similar. The strictest limit is obtained for the parametrized energy spectrum of Ref.~\cite{IT1996} as it predicts the hardest spectrum at high energy (see Fig.~\ref{fig:FluxComp}).

\section{Conclusion and discussion}
\label{sec:Conclusion}
We have performed the first experimental search for SA$\nu$ using data collected by the IceCube Neutrino Observatory during a 7 year period for the austral winter season when the declination of the Sun is above -5\degree. An unbinned likelihood analysis was performed with a total analysis livetime of 1406.62~days but no evidence for SA$\nu$s was found in the experimental data. The experimental data show an under-fluctuation relative to the background prediction and are consistent with a statistical fluctuation in the data. After inclusion of systematic uncertainties on the background prediction and signal efficiency, a 90\% confidence level upper limit is placed on the SA$\nu$ flux at 1~\textrm{TeV} of $1.02^{+0.20}_{-0.18}\cdot10^{-13}$~$\mathrm{GeV^{-1}cm^{-2}s^{-1}}$ for the benchmark signal energy spectrum from Ref.~\cite{Edsjo:2017kjk}. At present, our limit is about a factor of 13 larger than the baseline signal expectation. The results presented in this paper do not allow us to distinguish between various model predictions. In the future several improvements can be expected that will result in a better sensitivity to solar atmospheric neutrinos. The IceCube Upgrade will increase the acceptance of neutrinos down to a few GeV and a comprehensive calibration campaign is expected to reduce uncertainties related to ice properties, which will reduce reconstruction uncertainties in energy and arrival direction. Next-generation neutrino observatories, such as IceCube-Gen2~\cite{Aartsen:2020fgd} or KM3NeT~\cite{Adrian-Martinez:2016fdl}, will significantly increase acceptance of multi-TeV events, which are expected to have low atmospheric neutrino backgrounds. They could provide sufficient sensitivity to find evidence of SA$\nu$s.

The SA$\nu$ production mechanism is closely related to that of gamma rays from the solar disk. Based on public Fermi data it has been shown that the gamma-ray flux from the solar disk is about one order of magnitude higher than predicted~\cite{Seckel1991} and that the flux shows a significant time variation that anti-correlates with solar activity~\cite{Kenny2016,Tim2018,Qing2018}. We point out that our IceCube dataset, which consists of all available data at the time of the analysis, covers the period from May 2010 till May 2017 and hence does not include the solar minimum. Similar to the observed flux increase in gamma rays, the SA$\nu$ flux is expected to be enhanced during the solar minimum~\cite{Masip:2017gvw}. Based on the current data and models predictions, we point out that there is a considerable uncertainty in the flux expectations of solar gamma-rays and neutrinos. In addition, energetic gamma-ray events are observed primarily during the solar minimum. A continuation of this analysis during the solar minimum of 2019-2020 is in progress.

An observation of solar atmospheric neutrinos would be important to understand solar atmospheric magnetic fields, cosmic ray interactions in the solar atmosphere, and cosmic ray propagation in the inner solar system. Furthermore, a measurement of the SA$\nu$ flux is also essential for solar dark matter searches to characterize the SA$\nu$ sensitivity floor~\cite{Kenny2017, FJAWs2017,Edsjo:2017kjk}. If the SA$\nu$ flux is experimentally measured, it will provide the normalization of this irreducible background for solar dark matter searches.

Given the expected sensitivity of this analysis we decided to only test one signal hypothesis, namely our baseline model. In future analyses with improved sensitivity, differential flux limits could be produced and a more model independent approach could be taken to probe different energy ranges.

Lastly, an observation of the SA$\nu$s can be exploited as a calibration source for neutrino telescopes in the future. An observation of a high-energy neutrino signal from the Sun would only be the second of its kind, following the recent evidence of a high-energy neutrino signal from the blazar TXS 0506+056~\cite{IC2018_TXSA,IC2018_TXSB}. 

\acknowledgments
The IceCube collaboration acknowledges the
significant contributions to this manuscript from Seongjin In and Carsten Rott.
The authors gratefully acknowledge the support from the following agencies and institutions:
USA {\textendash} U.S. National Science Foundation-Office of Polar Programs,
U.S. National Science Foundation-Physics Division,
Wisconsin Alumni Research Foundation,
Center for High Throughput Computing (CHTC) at the University of Wisconsin-Madison,
Open Science Grid (OSG),
Extreme Science and Engineering Discovery Environment (XSEDE),
U.S. Department of Energy-National Energy Research Scientific Computing Center,
Particle astrophysics research computing center at the University of Maryland,
Institute for Cyber-Enabled Research at Michigan State University,
and Astroparticle physics computational facility at Marquette University;
Belgium {\textendash} Funds for Scientific Research (FRS-FNRS and FWO),
FWO Odysseus and Big Science programmes,
and Belgian Federal Science Policy Office (Belspo);
Germany {\textendash} Bundesministerium f{\"u}r Bildung und Forschung (BMBF),
Deutsche Forschungsgemeinschaft (DFG),
Helmholtz Alliance for Astroparticle Physics (HAP),
Initiative and Networking Fund of the Helmholtz Association,
Deutsches Elektronen Synchrotron (DESY),
and High Performance Computing cluster of the RWTH Aachen;
Sweden {\textendash} Swedish Research Council,
Swedish Polar Research Secretariat,
Swedish National Infrastructure for Computing (SNIC),
and Knut and Alice Wallenberg Foundation;
Australia {\textendash} Australian Research Council;
Canada {\textendash} Natural Sciences and Engineering Research Council of Canada,
Calcul Qu{\'e}bec, Compute Ontario, Canada Foundation for Innovation, WestGrid, and Compute Canada;
Denmark {\textendash} Villum Fonden, Danish National Research Foundation (DNRF), Carlsberg Foundation;
New Zealand {\textendash} Marsden Fund;
Japan {\textendash} Japan Society for Promotion of Science (JSPS)
and Institute for Global Prominent Research (IGPR) of Chiba University;
Korea {\textendash} National Research Foundation of Korea (NRF);
Switzerland {\textendash} Swiss National Science Foundation (SNSF);
United Kingdom {\textendash} Department of Physics, University of Oxford.

\bibliography{bibliograpy}

\renewcommand{\arraystretch}{1.2}
\begin{sidewaystable}[t]
\begin{center}
\begin{tabular}{|l|c|c|c|c|c|c|}
\hline
Energy spectrum (RoI, 1406.62 days) & $\bar{n}_{\textrm{sig}}$ & $\mu_{90}$ & $\hat{\mu}$ (Data) & TS (Data) & p-value & $\Phi_{90\%}(1~\textrm{TeV})$ [$\mathrm{GeV^{-1}cm^{-2}s^{-1}}$] \\
\hline
\hline
\multicolumn{7}{|l|}{Baseline model: Ingelman \& Thunman (1996)~\cite{IT1996}} \\
\hline
IT Flux (1996) & 2.83 & 35.09 & -2.28 & -0.05 & 0.56 & $8.57\cdot10^{-14}$ \\
\hline
\hline
\multicolumn{7}{|l|}{Reference model 1: Edsj\"{o} et al. (2017)}~\cite{Edsjo:2017kjk} \\
\hline
Serenelli~\cite{Serenelli}-GS98~\cite{GS98}-H3a~\cite{H3A}-Normal & 2.85 & 36.56 & -1.95 & -0.03 & 0.55 & $1.01\cdot10^{-13}$ \\
\hline
Serenelli-Stein~\cite{Stein}-H3a-Normal (Baseline) & 2.80 & 36.52 & -1.96 & -0.03 & 0.55 & $1.02\cdot10^{-13}$ \\
\hline
Serenelli-GS98~\cite{GS98}-H3a-Inverted & 2.95 & 36.26 & -1.93 & -0.03 & 0.55 & $9.65\cdot10^{-14}$ \\
\hline
Serenelli-Stein-H3a-Inverted & 2.89 & 36.39 & -1.96 & -0.03 & 0.55 & $9.89\cdot10^{-14}$ \\
\hline
Serenelli-GS98-4Gen~\cite{GST4Gen}-Normal & 2.70 & 37.21 & -1.93 & -0.03 & 0.55 & $1.08\cdot10^{-13}$ \\
\hline
Serenelli-Stein-4Gen-Normal & 2.65 & 37.30 & -2.00 & -0.03 & 0.55 & $1.10\cdot10^{-13}$ \\
\hline
Serenelli-GS98-4Gen-Inverted & 2.79 & 36.98 & -1.96 & -0.03 & 0.55 & $1.04\cdot10^{-13}$ \\
\hline
Serenelli-Stein-4Gen-Inverted & 2.73 & 37.06 & -1.95 & -0.03 & 0.55 & $1.06\cdot10^{-13}$ \\
\hline
\hline
\multicolumn{7}{|l|}{Reference model 2: FJAWs (2017)~\cite{FJAWs2017}} \\
\hline
SIBYLL2.3-pp~\cite{SIBYLL_A,SIBYLL_B}-CombinedGH-H4a~\cite{CombineH4A} & 2.16 & 38.40 & -2.08 & -0.03 & 0.56 & $1.37\cdot10^{-13}$ \\
\hline
SIBYLL2.3-pp-GaisserHonda~\cite{GH} & 1.82 & 38.40 & -2.34 & -0.04 & 0.56 & $1.66\cdot10^{-13}$ \\
\hline
SIBYLL2.3-pp-HillasGaisser-H4a~\cite{H3A} & 2.17 & 37.51 & -2.07 & -0.03 & 0.56 & $1.36\cdot10^{-13}$ \\
\hline
SIBYLL2.3-pp-PolyGonato~\cite{Poly} & 1.74 & 38.32 & -2.19 & -0.03 & 0.56 & $1.73\cdot10^{-13}$ \\
\hline
SIBYLL2.3-pp-Thunman~\cite{Thunman} & 1.95 & 38.33 & -2.28 & -0.04 & 0.56 & $1.55\cdot10^{-13}$ \\
\hline
SIBYLL2.3-pp-ZatsepinSokolskaya~\cite{ZS} & 1.71 & 37.45 & -2.18 & -0.04 & 0.56 & $1.72\cdot10^{-13}$ \\
\hline
SIBYLL2.3-ppMRS~\cite{MRS}-CombinedGH-H4a & 2.17 & 37.53 & -2.09 & -0.03 & 0.56 & $1.36\cdot10^{-13}$ \\
\hline
SIBYLL2.3-ppMRS-GaisserHonda & 1.82 & 38.39 & -2.34 & -0.04 & 0.56 & $1.65\cdot10^{-13}$ \\
\hline
SIBYLL2.3-ppMRS-HillasGaisser-H4a & 2.17 & 37.44 & -2.08 & -0.03 & 0.56 & $1.35\cdot10^{-13}$ \\
\hline
SIBYLL2.3-ppMRS-PolyGonato & 1.75 & 38.22 & -2.22 & -0.04 & 0.56 & $1.72\cdot10^{-13}$ \\
\hline
SIBYLL2.3-ppMRS-Thunman & 1.95 & 38.24 & -2.29 & -0.04 & 0.56 & $1.54\cdot10^{-13}$ \\
\hline
SIBYLL2.3-ppMRS-ZatsepinSokolskaya & 1.71 & 37.50 & -2.21 & -0.04 & 0.56 & $1.72\cdot10^{-13}$ \\
\hline
\end{tabular}
\end{center}
\caption{\label{tab:Sensitivity}Summary table of the analysis results for corresponding models of the energy spectra from each reference in the first column. ``Normal'' and ``Inverted'' in the rows of Ref.~\cite{Edsjo:2017kjk} refer to the neutrino mass ordering. Columns 2-7 represent the expected number of signal events, $\bar{n}_{\textrm{sig}}$, the Poisson mean for the 90\% C.L. limit, $\mu_{90}$, the maximum likelihood estimator, $\hat{\mu}$, the observed TS value, p-value and the flux limit at 1~TeV, $\Phi_{90\%}(1~\textrm{TeV})$.}
\end{sidewaystable}

\end{document}